# Mitigating the ICA Attack against Rotation Based Transformation for Privacy Preserving Clustering

Abedelaziz Mohaisen and Dowon Hong

*ABSTRACT— The rotation based transformation (RBT) for privacy preserving data mining (PPDM) is vulnerable to the independent component analysis (ICA) attack. This paper introduces a modified multiple rotation based transformation (MRBT) technique for special mining applications mitigating the ICA attack while maintaining the advantages of the RBT.*

## I. Introduction

While it is important for data owners to publish their data to a third party for providing data mining services, the privacy of the data itself needs to be maintained. Therefore, several perturbation methods have been introduced considering potential applications. One of these methods is the rotation based transformation (RBT) in which the data is transformed geometrically while preserving the distance between the data points. Such distance-preservation is vital for providing a high accuracy that reflexes a minimal data loss when performing data clustering [1].

For privacy, perturbation schemes generally, and the RBT scheme specially, are studied in different methods considering attacker's ability and knowledge. These methods are: the naïve estimation-based attack, the reconstruction-based attack, and the distance inference-based attack [2]. While the first attacking method is infeasible for most of the perturbation methods and the second is not applicable to the RBT, the distance inference-based attack has some impact. This impact was studied recently where two statistical tools, namely the Principle Component Analysis (PCA) [3] and the Independent Component Analysis (ICA) [4], have been utilized. Both tools have shown a relevant efficiency in breaching the privacy under some operating conditions.

In this letter we revise the ICA attack on the RBT and introduce a multiple rotation-based transformation (MRBT) method that helps mitigating the ICA attack.

Technically, RBT is a method that has been introduced for data perturbation to guarantee an exact accuracy and maintain the privacy of geometrical and numerical data [2], [5]. The general transformation follows the model $Y=R_\theta X$ where $X$ is the original data, $R_\theta$ is a rotation matrix and $Y$ is the rotated (i.e., transformed) data to be released to a third party. The rotation matrix $R_\theta$ needs to be *Orthogonal* in order to satisfy the distance-invariant property. That is, $R_\theta R_\theta^T = R_\theta^T R_\theta = I$ where $I$ is the identity matrix. The RBT preserves the vector length, Euclidean distance, and inner product between two vectors which are essential in many clustering algorithms.

On the other hand, the ICA is a method for independent components separation that aims to separate two or more signals with some specific properties following the form $Y=AS$, where $S$ is the set of independent components (i.e., data) and $A$ is mixing matrix (corresponding to $X$ and $R$ in the RBT respectively). The ICA is subject to several restrictions on the used data including that the data attributes need to be linearly independent and with non-Gaussian distribution (except of one attribute at most).

The *A-priori Knowledge ICA* (AK-ICA) is a simple modification of the ICA and has recently been shown to be effective on the RBT [4]. The AK-ICA works as follows: given a small portion of the original data, its transformed image, and the transformed image of the whole data, it is then possible to recover the whole data (i.e., population) using the ICA given enough information about the distribution of the original private data. That is, the attacker firstly applies the ICA on the whole transformed dataset then applies it again on the known private data portion. From the separated components, he finds some matrix $J$ that maximizes the mutual information between the separated components by computing (at a point $z$)

$$I(f_i, f_j') = \tfrac{1}{2} E[\int_{\Omega z} |f_i(z) - f_j'(z)| dz]$$

Where $f_i$ and $f_j'$ are the density distributions of the i-th component and the j-th component of original dataset and the reconstructed dataset respectively. The smaller $I$, the more similar the estimated and original data are.

The power of the ICA attack is based on that the known fraction of data has sufficient information about the distribution of the whole data population. In practice, it is difficult to mount such attack by only observing a small portion of the data.

In this paper, we introduce a multiple RBT (MRBT) scheme that mitigates the impact of AK-ICA. The idea of the MRBT is based on dividing the whole data into sets and transforming them independently in order to harden the process of the recovering the ICA components and applying the AK-ICA. Experimentally, our scheme shows a reasonable efficiency in mitigating AK-ICA. We also show that AK-ICA attack on geometric data (in [7]) produces lower accuracy than the accuracy of the dataset in [4].

## II. Multiple RBT (MRBT) and Applications

The main goal of our rotation scheme is to preserve the distance and the inner product between data sets *partially*, as a valid goal for different applications, while providing a high privacy for the transformed data. The scheme is summarized in the following section where data is assumed to be numerical

a) The data owner normalizes the data to unity.
b) According to some chosen in-advance $n$, the data owner divides the data into $n$ equal parts defined as follows

$$X' = \{X_1' \| X_2' \| X_3' \| \cdots \| X_n'\}$$

c) The data owner generates $n$ different random seeds. Using each seed $i$, the data owner generates an *orthogonal* matrix $\mathbf{R}_{\theta i}$ for rotating the corresponding part of $\mathbf{X}$
d) The data owner transforms his data as follows:

$$Y' = \{Y_1' \| \cdots \| Y_n'\} = \{R_{\theta 1} X_1' \| \cdots \| R_{\theta n} X_n'\}$$

e) The data owner releases the rotated data for public use.

The resulting rotation preserves inner product between *corresponding tuples* in the original data. Also, it preserves the inner product between two tuples falling into *same corresponding subsets*. However, the inner product for other than the aforementioned tuples is not preserved. The first claim can be easily proven given that these tuples are transformed using the same matrix (follows from [1].) Similarly, we prove the second claim as follows. Consider the following rotated data sets:

$$Y' = \{Y_1' \| \cdots \| Y_n'\} = \{R_{\theta 1} X_1' \| \cdots \| R_{\theta n} X_n'\}$$

$$Y'' = \{Y_1'' \| \cdots \| Y_n''\} = \{R_{\theta 1} X_1'' \| \cdots \| R_{\theta n} X_n''\}$$

The inner product between these two datasets is

$$Y'^T Y'' = \begin{pmatrix} Y_1'^T Y_1'' & Y_1'^T Y_2'' & \cdots & Y_1'^T Y_n'' \\ Y_2'^T Y_1'' & Y_2'^T Y_2'' & \cdots & Y_2'^T Y_n'' \\ \vdots & \vdots & \ddots & \vdots \\ Y_n'^T Y_1'' & Y_n'^T Y_2'' & \cdots & Y_n'^T Y_n'' \end{pmatrix}$$

For the diagonal part of the above product matrix, it is easy to verify the preservation of the inner product.

$$Y_i'^T Y_i'' = (R_{\theta i} X_i')^T R_{\theta i} X_i'' = X_i'^T R_{\theta i}^T R_{\theta i} X_i''$$
$$= X_i'^T I X_i'' = X_i'^T X_i''$$

therefore, the second claim is proven by the above result. Also, the distance between the corresponding subsets can be easily driven using the following formula [3].

$$d(X'', X') = \sqrt{\sum_i (x_i'' - x_i')} = \sqrt{2 - 2 X''^T X'}$$

The quantification of the privacy in our scheme follows the same metrics in [4]. Furthermore, we use the maximization of differences' covariance between the original data and the rotated data [1]. The later metric can be systemically guaranteed by setting the different rotation matrices that maximize the difference covariance per data subset.

In our MRBT scheme, the geometrical shape of the data is distorted into $n$ different parts where a simple clustering algorithm will not work correctly. In lieu, different promising applications based on the coordinated pair-wise distance still exist where our scheme can be used for such applications.

*Application 1*: The first application considers computing the inner product of two vectors transformed using MRBT. This application is straightforward of the above scheme (on the data subsets). After releasing the transformed data, the third party computes the inner product on the corresponding subsets.

*Application 2*: The main concern is to compute the distance between the corresponding tuples of two private data vectors. Firstly, the two parties perform the routine in *Application 1* to obtain the inner product. After that, the third party plugs the resulting inner product in the distance/inner product formula in order to compute the distance..

*Application 3*: (coordinated distance in network diagnosis) A data owner has a real data representing site access log from which he would like to diagnose his own site by clustering different measured values (e.g., processing, bandwidth, etc) measured over the time slots at which an anomaly access has probably occurred at the site by comparing the difference of the access logs from some ideal case's log file. This application is directly transferred into a clustering problem over the resulting distance in accordance with the as an index. To do so, the following procedure is performed

a) The two sites generate a rotated version of their own log data using the method in *application 1* and release the rotated data to the third party.
b) The third party computes difference of access log for the corresponding days resulting in the matrix of differences.
c) The third party performs the clustering algorithm (e.g., k-mean) on the resulting set of distances.

Note that the matrix of differences is preserved regardless to the method of rotation. That is, though the single day's access log data is rotated using different angels, the computed difference is maintained same for the single day.

In addition to these applications, it easy to recall and extend other applications from the literature such like those in [8] and [9] in which the privacy of clustering (data or location information) can be of a great benefit.

## III. Impact of AK-ICA on the MRBT

To study the impact of the AK-ICA on the MRBT, we perform experiments for reconstructing the original data by observing the rotated data and the known original data fraction to the attacker. We use the same dataset used in [4].

*Experiment 1:* The used data is with normal distribution. A small fraction of it (i.e., 10%) can, using the Gaussian kernel density estimation, extrapolate the distribution of the whole data with high accuracy (as high as 94%)

*Experiment 2:* The accuracy of the reconstructed data is estimated according to the reconstruction error measure by the Frobenius norm. Figure 1 shows the reconstruction accuracy versus the fraction of known data to the attacker for different $n$. As we assume the boundaries of data (i.e., min and max) are known to the attacker, he can rescale the reconstructed data and solve the ambiguity of reconstruction amplitude noted in [2].

*Experiment 3*: Using UJI Pen Characters dataset [7], which represents a more realistic geographical data, we obtain an accuracy of 45.7% when applying AK-ICA attack for 10% known private data fraction and as high as 89.2% for the same data portion. For more details on additional experiments and comparisons, please see [6].

## IV. Conclusion

In this paper we introduced a scheme for reducing the impact of the ICA attack on the rotation based transformation which is commonly used for privacy preserving data clustering. Our results show a relative efficiency in reducing the ICA impact while introducing chances for several applications based on maintaining the coordinated distance and inner product between the corresponding subsets of transformed data.

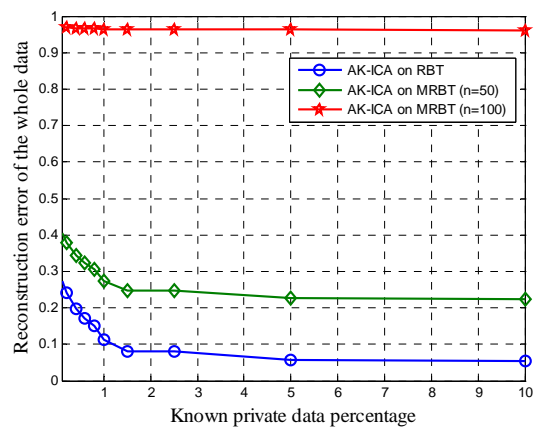

Figure 1. Reconstruction Error of MRBT with different $n$

## References

[1] K. Chen, L. Liu. "Privacy preserving data classification with rotation perturbation", *Proc. ICDM* '05, 2005, p. 589-592.
[2] K. Chen, G. Sun, L. Liu: "Towards Attack-Resilient Geometric Data Perturbation", *Proc. SDM* '07, p. 89-94
[3] K. Liu, H. Kargupta, and J. Ryan. "Random projection-based multiplicative data perturbation for privacy preserving distributed data mining", *IEEE Trans. Knowl. Data Eng.*, vol. 18 no. 1 Dec. 2006, p. 92-106.
[4] S. Guo, X. Wu: "Deriving Private Information from Arbitrarily Projected Data", *Proc. PAKDD* '07, p. 84-95.
[5] S. R. M. Oliveira and O. R. Zaiane, "Privacy preservation when sharing data for clustering". *Proc. SDM* '04, 2004, p. 67-82.
[6] A. Mohaisen, D. Hong: "Mitigating the ICA Attack against Rotation Based Transformation for Privacy Preserving Clustering", *technical report*, ETRI, Korea, March 2008. http://www.mohaisen.net/pdf/mrbttr/.
[7] UJI Pen Characters dataset, UCI ML datasets online: http://archive.ics.uci.edu/ml/datasets/UJI+Pen+Characters
[8] T. Nhan Vu, J. Lee, and K. Ryu, Spatiotemporal Pattern Mining Technique for Location-Based Service System, ETRI Journal, vol.30, no.3, June 09,, pp.421-431.
[9] S. Kang, D. Kim, Y. Lee, S. Hyun, D. Lee, and B. Lee, A Semantic Service Discovery Network for Large-Scale Ubiquitous Computing Environments, ETRI Journal, vol.29, no.5, Oct. 09,, pp.545-558.